\documentclass[12pt]{article}
\usepackage{color}
\usepackage{graphicx}
\usepackage{times}

\usepackage{scicite}

\newenvironment{sciabstract}{%
\begin{quote} \bf}
{\end{quote}}

\usepackage{amssymb}
\usepackage{amsmath}
\usepackage[utf8]{inputenc}

\DeclareMathOperator{\sinc}{sinc}
\DeclareMathOperator{\sign}{sign}

\newcommand{\degree}{$^\circ$}
\newcommand{\x}{\mathbf{x}}
\newcommand{\uu}{\mathbf{u}}

\usepackage{lineno}

\date{}

\begin{document}
\baselineskip24pt


\title{
Asymptotic Solution to the Rayleigh Problem of Dynamic Soaring\\
}
\author{Gabriel D. {Bousquet},$^{1\ast}$ Michael S. Triantafyllou,$^{1}$ Jean-Jacques E. Slotine$^{1}$}

\maketitle

$^{1}$Department of Mechanical Engineering, Massachusetts Institute of Technology, Cambridge, MA 02139,  USA.\\
 $^\ast$Corresponding author. E-mail: g\_b@mit.edu

\modulolinenumbers[5]
\linenumbers
\noindent\begin{sciabstract}
Albatrosses can travel a thousand kilometers daily over the oceans. This feat is achieved through dynamic soaring, a non-flapping flight strategy where propulsive energy is extracted from horizontal wind shears. Dynamic soaring has been described as a sequence of half-turns connecting upwind climbs and downwind dives through the surface shear layer. We analytically and numerically investigate the aerodynamically optimal flight trajectory for varying shear thicknesses. Contrary to current thinking, but consistent with \textsc{gps} recordings of flying albatrosses, in thin shears the optimal trajectory is composed of small angle arcs. Essentially, the albatross is a flying sailboat, sequentially acting as sail and keel, and most efficient when remaining crosswind. Our analysis constitutes a general framework for dynamic soaring, and more broadly energy extraction in complex winds.
\end{sciabstract}

\noindent Dynamic soaring (DS) is the flight technique where a glider, either a bird or man-made, extracts its propulsive energy from non-uniform  horizontal winds such as those found over the oceans~\cite{Rayleigh1883}. Albatrosses, the archetypal dynamic soarers, have been recorded to travel 5000~km per week  while relying on wind energy alone~\cite{Sachs2012,Catry2004,Weimerskirch2000}. 
The potentialities of man-made DS systems are humbling: a robotic albatross could patrol the oceans and collect oceanic and atmospheric data, traveling at over 40~knots with a virtually infinite range~\cite{Bower2011,Langelaan2009a}. Another flavor of dynamic soaring, harvesting the wind shear of the jet stream, is an active field of research~\cite{Sachs2006}.

Despite important efforts~\cite{Deittert2009,Zhao2004,Boslough2002,Barate2006,Bower2011,Lawrance2011,Bonnin2013,Bird2014,Gao2015}, a DS robotic system has remained out of reach. A major obstacle to man-made DS has resided in the complexity of the wind power extraction process that, by nature, requires planning on the go an energy positive trajectory in a stochastic, hard to measure, and poorly understood wind field. 
Until now, the fundamentals of the energetic exchange in DS have only been partially understood, preventing us from efficiently solving the planning problem.
Here, we provide a surprising and yet very intuitive description of DS, backed up by analytic and numerical analysis as well as comparison with field data.

\begin{figure}[p!]
  \centering
  \includegraphics[width=2.9in]{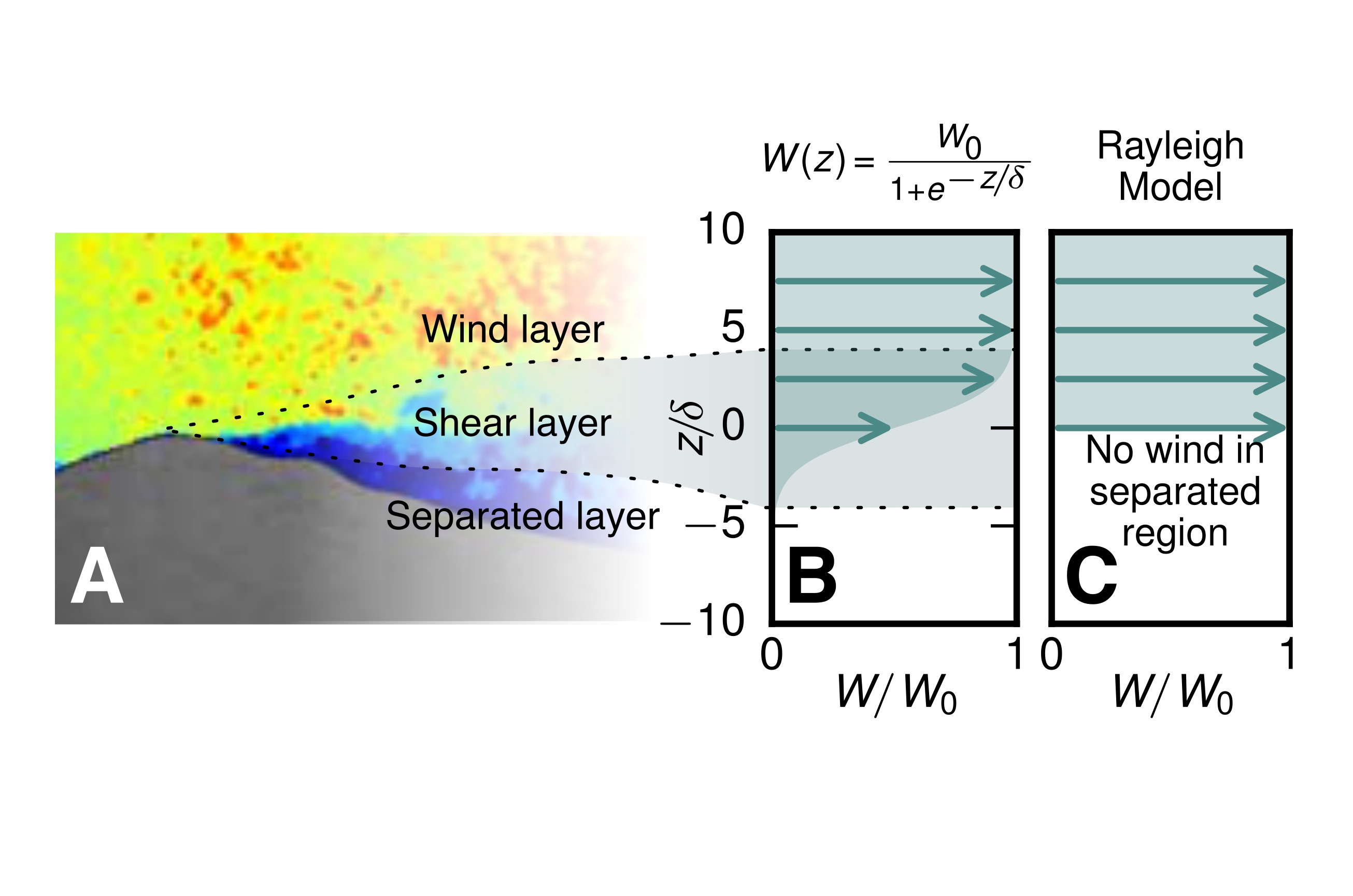}
  \caption{\textbf{Wind profile}.  (\textbf{A}) Wind field behind waves. Color-coding: wind intensity. Experimental data adapted from \cite{Veron2007}. (\textbf{B}) The logistic wind profile in this study captures adequately the wind field in separated regions, such as behind ocean waves. More generally, it constitutes a robust way to approximate a wide class of wind fields, based on two parameters: a typical wind speed inhomogeneity $W_0$ separated by a buffer of typical dimension $\delta$. (\textbf{C}) Rayleigh's wind model is the limit of the logistic profile for $\delta\to 0$.}
  \label{fig:veron2007}
\end{figure}

\begin{figure}
  \centering
  \includegraphics[width=\textwidth]{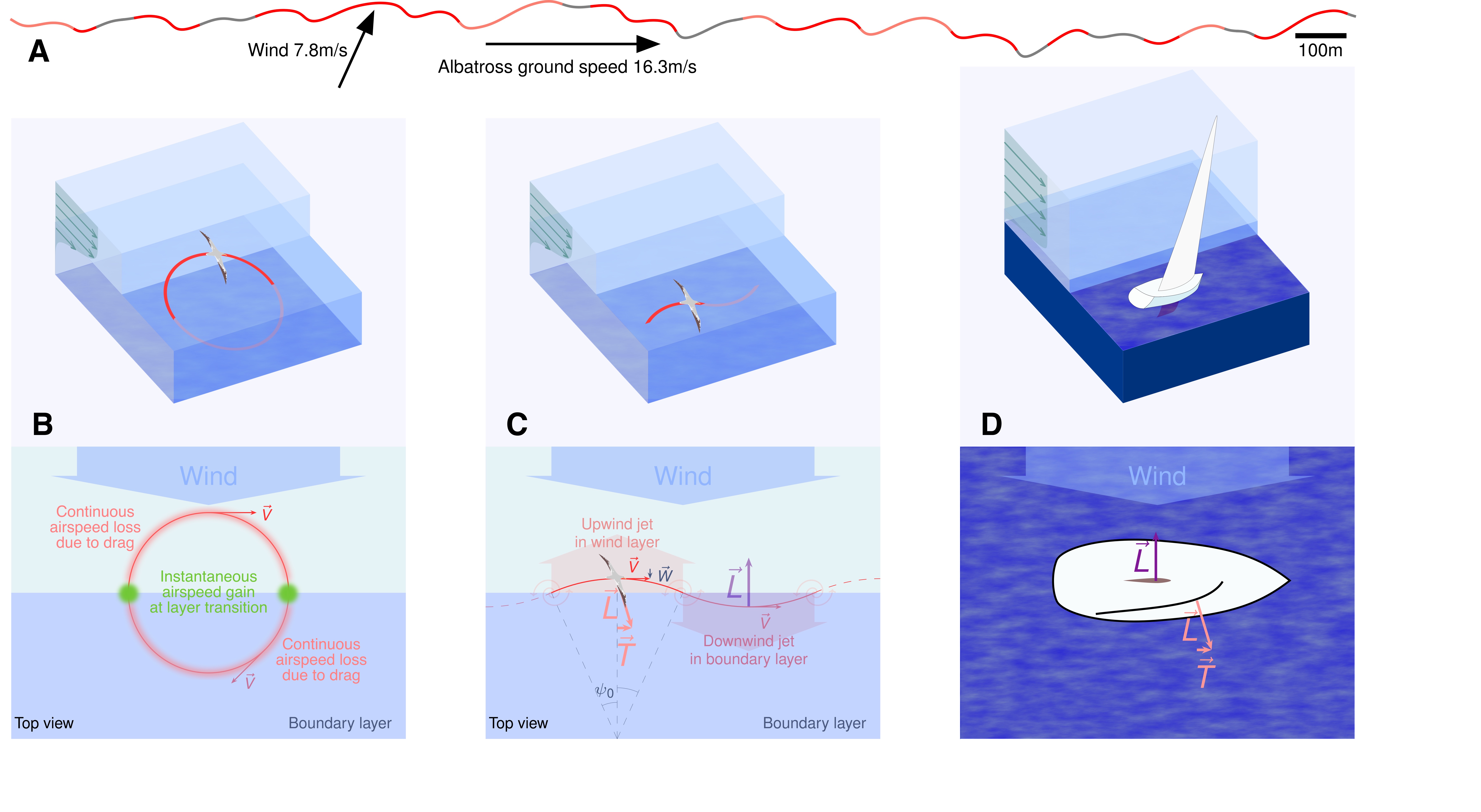}
  \caption{  \label{fig:synthesis}\textbf{The albatross' trajectory}. 
(\textbf{A}) Recording of a flying albatross from~\cite{Sachs2016} (top view). In crosswind flight the typical turn of the albatross is less than 60\degree{} (\emph{c.f.} ST4). Orange portions of the trajectory: the albatross is involved in a $60\pm20^\circ$ turn. Red portions: the albatross is involved in a $60\pm10^\circ$ turn. 
Note that while in the ground frame the mean albatross travel has a downwind component, in the frame moving with the average wind it is nearly crosswind.
(\textbf{B}) The Rayleigh cycle describes the albatross' flight as a sequence of half-turns between the windy and slow regions. At each layer transition, there is an airspeed gain equal to the wind speed, which compensates inherent drag losses that are quadratic in airspeed. 
However this trajectory is suboptimal for energy extraction. Instead, the optimal cycle (\textbf{C}) is composed of a succession of small angle arcs. The flight portion in the wind layer is functionally analogous to the sail of a sailboat while the portion in the slow layer is analogous to the keel of a sailboat (\textbf{D}).}
\end{figure}

In the first attempt to describe DS, Rayleigh~\cite{Rayleigh1883} modeled the wind profile as a still boundary layer separated from the above windy free stream blowing at $W_0$  by an infinitely thin shear layer (see Fig.~\ref{fig:veron2007}C, hereafter Rayleigh's wind model).
He noticed that when traversing the shear layer directly up- or downwind, the albatross' groundspeed is conserved but its airspeed is not, and may increase by up to $W_0$. Rayleigh connected up- and downwind transitions with half-turns in order to construct an energy neutral trajectory (hereafter Rayleigh's cycle, Fig.~\ref{fig:synthesis}B and \emph{e.g.} \cite{Richardson2011}): at each transition, the airspeed gain compensates the inherent losses due to drag. Because the drag is quadratic with airspeed, a limit cycle is reached. This description of the DS trajectory has carried on until today~\cite{Pennycuick1982,Boslough2002,Lissaman2005,Barate2006,Denny2009,Richardson2011,Bower2011,Lawrance2011,Bonnin2013,Bird2014,Gao2015} in two energetically equivalent forms: trajectories with constant turn direction are O-shaped, or loitering; trajectories with alternating turn directions are S-shaped, or traveling.

Recent observations based on high-accuracy \textsc{gps} measurements~\cite{Sachs2012,Sachs2016} (reproduced in Figs.~\ref{fig:synthesis}A and \ref{fig:sachs2016}) show that albatrosses do not follow  half-turns, but rather an elongated, albeit oscillating, trajectory. Statistical analysis of this data shows that
when flying crosswind albatrosses typically only turn by about 55$^\circ$, a mere third the half-turn's 180$^\circ$. In this article we explain how fundamental this distinction is. We numerically studied the dependence of the aerodynamically optimal trajectory of DS on shear layer thickness.
We discovered that contrary to prevailing theory, in the thin shear layer regime it is a sequence of arcs of vanishingly small angle, with the direction of flight nearly crosswind at all times. We were able to explain this observation analytically, lowering the wind required for DS by over 35\% compared to previous models~\cite{Bower2011}. 
For the albatross, the half-turn picture with up- and downwind transitions is misleading, as it is suboptimal both energetically and for travel speed. 
Our theory conceptually unifies dynamic soaring, gust soaring~\cite{Pennycuick2002}, turbulence soaring~\cite{Phillips1975,Patel2008}, and other wind energy harvesting techniques such as sailing.

The energy extraction mechanism in DS relies on a transfer of momentum from fast to slow air, and any theory starts with formulating the structure of the wind field.
In the last two decades, a popular approach has consisted in attempting to perform accurate numerical modeling of the albatross flight 
in logarithmic or power law profiles, deemed good models of the average wind field in the first 20~m above water
, where the albatross flies. However, in this framework it has been shown~\cite{Flanzer2012,Deittert2009} that DS is extremely sensitive to the wind field in the first meter above the surface, precisely where wind-wave interactions and temporal variability make the logarithmic model less accurate.

In contrast, Rayleigh's wind model has merit beyond the realm of qualitative analysis for modeling the sharp wind shear in separated regions, such as behind breaking waves or mountain ridges. Recent studies suggest that wind separation in ocean wave fields may be more frequent than previously believed (\cite{Veron2007,Buckley2016,Banner1976} and Fig.~\ref{fig:veron2007}A), further reducing the relative merit of log-based approaches.

 We modeled the wind with a logistic profile (Fig.~\ref{fig:veron2007}B) 
parameterized by the free stream wind speed $W_0$ and the shear layer thickness $\delta$ 
\begin{equation}
W({z}) = \frac{W_0}{1 + \exp{-{z}/\delta}}.\label{eq:wind}
\end{equation}
This captures not only the main features of separated winds over ocean waves but more generally of any flow with a typical wind inhomogeneity $W_0$ developing over a typical length-scale $\delta$, such as in turbulence soaring.
The regions $z\ll -\delta$, $|z|\lesssim 4\delta$, $z\gg\delta$ represent the boundary layer or separated region, shear layer and windy free stream layer, respectively. In the thin shear limit $\delta\to 0$ the model converges to  Rayleigh's.

We numerically addressed the following question: for a given shear thickness $\delta$, what is the trajectory that requires the least amount of wind? We applied a direct collocation approach to a 3D point mass glider model defined by its wing loading $m/S$ and lift-drag coefficient curve $c_L\mapsto c_D(c_L)$, in particular its glide ratio $(c_L/c_D)_{\max}$ (\emph{c.f.} ST1 and \cite{Zhao2004,Bower2011,Flanzer2012}). The model non-dimensionalization shows the importance of two  parameters:  the glider's cruise speed  $V_c = \sqrt{\frac{mg}{\frac{1}{2}\rho S}}$ and its associated length $\lambda = V_c^2/g$. Typical values for the wandering albatross are $V_c = 15$~m/s and $\lambda = 22$~m \cite{Sachs2005}.
\begin{figure}
  \centering
  \includegraphics{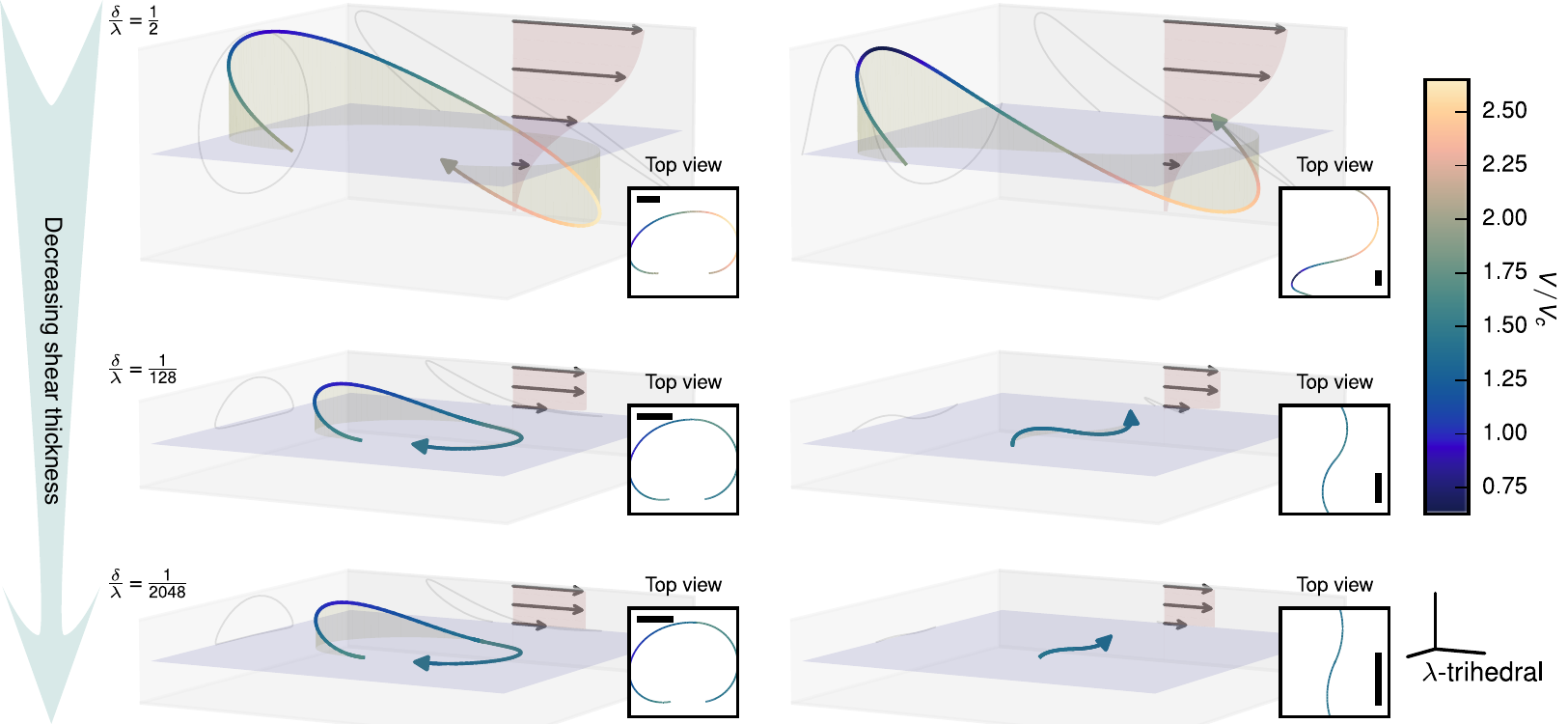}
  \caption{\textbf{Minimum wind trajectories} for three shear layer thicknesses (\emph{c.f.} ST2). On the left, the trajectories are constrained to fulfill the specific requirement that the heading increases by 360\degree{}  over a cycle, hence their loitering appearance. On the right, the heading is required to be periodic, hence their traveling appearance. 
 For the 3D trajectory the scale is common and is indicated on the bottom right corner: the trihedral is of length $\lambda= \frac{2m}{\rho S}$ (22~m for an albatross). Similarly, the scale bars on the top views are of length $\lambda$.  The middle plots $\delta/\lambda=1/128$ are representative of the shear thickness experienced by albatrosses. 
The traveling trajectory requires less wind than the loitering one, with an increasing advantage for thinner shears.
When $\delta\to 0$, the traveling trajectory becomes 2D and is composed of a sequence of vanishingly small arcs of finite curvature performed at nearly constant speed. The behavior of the loitering trajectory is qualitatively different: for decreasing shear thicknesses, it quickly converges to a limit trajectory that remains significantly 3D even for an infinitely thin shear layer.}
  \label{fig:numerical}
\end{figure}

Starting from thick shear ($\delta\gg \lambda$) and reducing it progressively until the thin shear regime ($\delta\ll \lambda$) was reached, we computed the optimal loitering (circling and therefore constrained to be half-turn based) and traveling trajectories (Fig.~\ref{fig:numerical} and ST2). 

\begin{figure}[p!]
  \centering
  \includegraphics{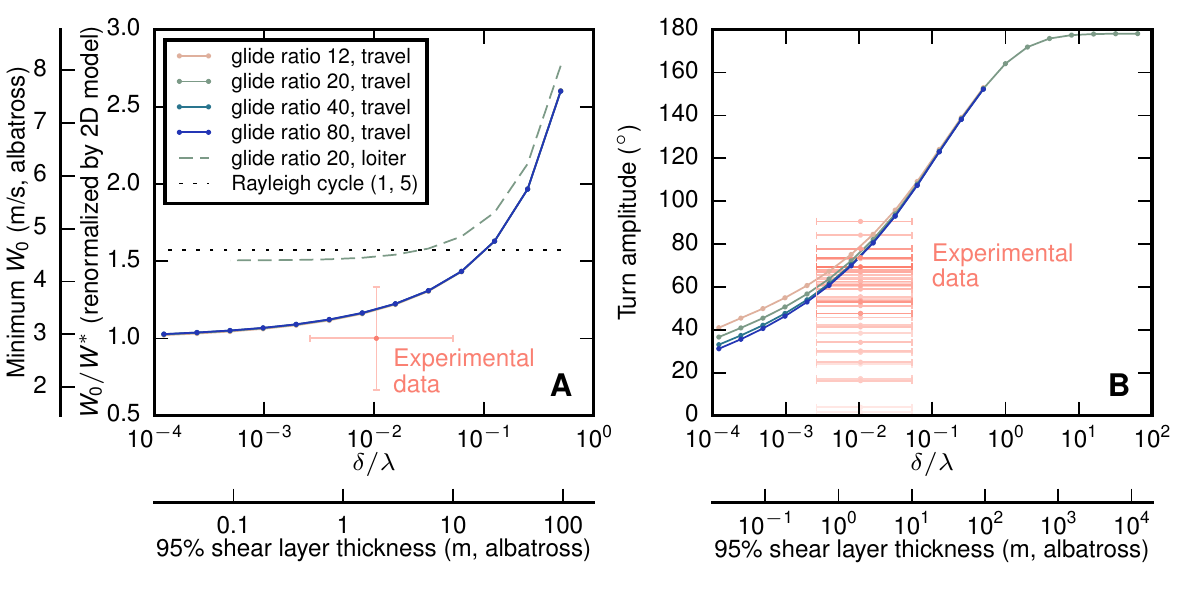}
  \caption{\textbf{Minimum wind and turn amplitude} of the traveling and loitering trajectories as a function of the shear thickness from our numerical model (\emph{c.f.} ST2).
(\textbf{A}) For thin shear $\delta\to 0$ the wind required for the traveling trajectories converges to our 2D model in Eq.~(\ref{eq:DSopt}). Note that all glide ratios collapse on nearly the same line. (\textbf{B}) Similarly, the turn amplitude decreases and the trajectories become straighter. The  data from~\cite{Sachs2016} is overlaid: each orange line on the right graph represents a turn from the albatross recording in Fig.~\ref{fig:synthesis}A (\emph{c.f.} ST3 and ST4).}
  \label{fig:2dpredict}
\end{figure}

In thick shear all trajectories are significantly three-dimensional, the loitering and traveling trajectories are quantitatively similar, and the turn amplitude of the traveling trajectory is large.
In thin shear, the loitering and traveling trajectories are qualitatively different. While the loitering trajectory remains significantly 3D, the traveling trajectory's extension in the $z$-direction shrinks and it becomes approximately 2D. For very thin shears it approaches a quasi-straight line, composed of zigzags of only a few degrees in amplitude. Most importantly, it requires only about 2/3 as much wind as the loitering trajectory (Fig.~\ref{fig:2dpredict}). The main characteristics of our numerical model  are strikingly consistent with albatross flight data, especially given the uncertainty associated with the wind field.

The convergence of the traveling trajectory to the neighborhood of $z = 0$ greatly reduces the problem complexity and it was possible to build an analytic model in this limit (Fig.~\ref{fig:synthesis}B and ST1). The cycle may be decomposed into glide phases on either side of, but close to, $z = 0$ where the wind shear is weak and  airspeed is lost due to drag, and transitions across $z = 0$ of vanishing duration but finite impulse (called ``swoops'' in \emph{e.g.} \cite{Pennycuick2002}). Denote $\psi$ the air-relative heading angle, defined to be $0$ when the glider is flying crosswind. During glide the 3D equations under 2D constraint simplify to $dV/d|\psi| = -\frac{c_D}{c_L}\frac{V}{\sqrt{1 - V_c^4/c_L^2V^4}}$. During the transition through $z = 0$, for large glide ratios the change in airspeed is $\Delta V = W_0 \sin|\psi_0|$. Balancing the airspeed loss during glides and airspeed gain at transitions brings the relation between wind intensity  $W_0$, average airspeed $V$ and heading angle at transition $\psi_0$:
\begin{equation}
\frac{V}{\sqrt{1 - V_c^4/c_L^2 V^4}} = \frac{\sin\psi_0}{\psi_0}\frac{c_L}{2c_D}W_0.
\label{eq:DSfundamental}
\end{equation}
The minimum wind-airspeed pair
\begin{equation}
W^* = \frac{3^{3/4}\sqrt{2}}{c_L^{3/2}/c_D}V_c,\quad V^* = \frac{3^{1/4}}{\sqrt{c_L}}V_c\label{eq:DSopt}
\end{equation}
is reached for $\psi_0\to 0$ (small amplitude turns), \emph{not} for $\psi_0 = \pi/2$ (half-turns). 
This represents the DS analogous to the famous Betz limit of wind energy~\cite{Manwell2009}.
Furthermore, it can be shown that it constitutes the benchmark for soaring in transverse turbulence when the characteristic turbulence scale is small compared to $\lambda$. Lastly, our analytic model is consistent with our numerical results (Fig.~\ref{fig:2dpredict}), and consequently provides strong qualitative agreement with the albatross trajectory recordings.

The previous discussion was based on airspeed considerations. With the understanding gained, we now discuss DS from an energetic standpoint in the ground frame of reference (Fig.~\ref{fig:synthesis}C). Power is extracted from the wind through the work done by the lift vector and is largest when the glider is flying close to crosswind. Indeed, the lift vector is always orthogonal to airspeed and flying crosswind allows for a large misalignment between airspeed and groundspeed, which in turns translates into a large dot product between the lift vector and groundspeed. 

The energetic exchange between the air and glider can be further understood by noticing that at each transition, the glider lift changes sign. 
By Kelvin's circulation theorem, vorticity must be shed at each transition.
Those vortices constitute the signature of jets formed in the wake of the glider which manifests a transfer of momentum: the passage of the glider slows down the wind layer and accelerates the boundary layer. That transfer of momentum is linked to an overall reduction of kinetic energy in the flow as it is harvested by the glider.

Thus, DS actually presents strong similarities to sailing:
sailboats transfer momentum from the fast wind to the slow moving ocean by means of two lifting surfaces, the sail and the keel, in a manner that is most efficient when operating near crosswind. A DS system effects a similar transfer of momentum from the fast wind to the slow boundary layer, playing \emph{sequentially} the roles of a sail while in the free stream and that of a keel while in the boundary layer (Fig.~\ref{fig:synthesis}C and D). Like the sailboat, it is most efficient when flying near crosswind. 

In practice, several factors limit the cycle frequency of DS. Our analytic model is strictly valid for an infinitely large glide ratio in an infinitesimal shear layer. For the albatross, those are finite: $c_L/c_D \lesssim 20$ and $\delta \sim 1$~m.
Furthermore, the large albatross wingspan ($\sim 3$~m) incurs a cost of rolling not taken into account in our point mass model, as well as a constraint on its minimum vertical travel. Finally, there may be other advantages to lower frequency cycles, such as the possibility to synchronize with  waves, and a reduction of the requirement in control authority. For the albatross, finite turns are not the cause of energy extraction, but a consequence of these practical limitations.

This new conceptual framework has important implications: beside constituting the basis for rich allometric extensions, the improved understanding of the albatross flight and the uncovering of the potentially major role played by wind separation behind waves in its soaring ability help refine the characterization and prediction of the albatross' habitat in a changing climate~\cite{Weimerskirch2012}. In the quest for a robotic, bioinspired albatross, Eq.~(\ref{eq:DSopt}) may well constitute the fundamental design guideline, while understanding the key roles of shear thickness and turn amplitude paves the way to robust and scalable learning algorithms.

\nolinenumbers

\bibliography{library}

\noindent\textbf{Acknowledgements:}
This research was supported in part by
a Fulbright Science and Technology fellowship, a Professor Amar G. Bose research grant, a Link Foundation Ocean Engineering \& Instrumentation fellowship as well as from \textsc{smart}, the Singapore-\textsc{mit} Alliance for Research and Technology, within the \textsc{censam} program.
We thank P. Beynet, L. Deike, E. Ferenczi, J. Izraelevitz, J. Schulmeister and A. Teferra for stimulating discussions and feedback.

\newpage
\linenumbers[1]

\section*{ST1 Analytic model}
\subsection*{Equations of motion and non-dimensionalization}

The analysis utilizes a 3-degree of freedom glider model. Our formulation follows closely \cite{Zhao2004,Bower2011} in the frame or reference $(\mathbf{i}, \mathbf{j}, \mathbf{k}) = (\mathbf{e}_\text{East},\mathbf{e}_\text{North},\mathbf{e}_\text{Up})$ with the state $\x = (V, \psi, \gamma, z,x,y)$ where $V$ is the glider airspeed, $\psi$ is the angle between $\mathbf{{x}}$ and the projection of the airspeed $\mathbf{V}$ in the $xOy$ plane and $\gamma$ is the angle between $\mathbf{V}$ and the $xOy$ plane and is positive nose up. $\psi$ and $\gamma$ are the air-relative heading angle and  air-relative flight path angle respectively. We assume the existence of a varying wind $-W(z)\mathbf{j}$ (blowing from positive to negative $y$ when $W>0$). The control inputs are the lift coefficient and bank angle $\uu = (c_L, \phi)$. The equations of motion (\textsc{eom}) are:

\begin{subequations}\label{eq:EOM}
  \begin{align}
    m \dot{V} &= -D - mg\sin\gamma + m\dot{W}\cos\gamma\sin\psi
\label{eq:xzvdot}\\
mV\dot{\gamma} &= L\cos\phi
 - mg\cos\gamma - m\dot{W}\sin\gamma\sin\psi
\label{eq:xzgammadot}\\
    mV\dot\psi\cos\gamma &= L\sin\phi + m\dot{W}\cos\psi\label{eq:psidot}\\
\dot{z}&= V\sin\gamma\label{eq:zdot}\\
\dot{x} &= V\cos\gamma\cos\psi\label{eq:xdot}\\
\dot{y} &=  V\cos\gamma\sin\psi  - W\label{eq:ydot}
  \end{align}
\end{subequations}
Note that $x,y$ may be considered as \emph{output} rather than \emph{states} as they don't feed back into Eqs.~(\ref{eq:xzvdot}--\ref{eq:zdot}). We are looking for the \emph{minimum wind} trajectories, \emph{i.e.} periodic in the states $V,\psi, \gamma, z$. In particular we do \emph{not} require staying upwind.
Lift and drag are specified according to  $L, D = 1/2c_{L, D}\rho S V^2$. We assume quadratic drag $c_D = c_{D,0} + k c_L^2$ with $k^{-1} = 4f_{\max}^2 c_{D, 0}$ where $f_{\max}$ is the glider maximum lift-to-drag ratio.

The problem is non-dimensionalized as follows: the natural velocity of the problem is the glider's cruise speed at $c_L =1$, namely $V_c = \sqrt{\frac{mg}{\frac{1}{2}\rho S}}$. The natural length is $\lambda = (V_c)^2/g$. The associated time-scale is $t_c = V_c/g = \lambda/V_c$. Note that our non-dimensionalization depends only on the glider properties and gravity,  arguably a more natural choice than approaches based on the wind gradient~\cite{Zhao2004}.

Upon non-dimensionalization of the variables $v = V/V_c$, $w = W/V_c$, $\tilde{z} = z/\lambda$, $\tau = t/t_c$ and $(\cdot)' = d(\cdot)/d\tau$, Eq.~(\ref{eq:EOM}) becomes

\begin{subequations}\label{eq:eom}
  \begin{align}
    v' &= -c_Dv^2 - \sin\gamma + w'\cos\gamma\sin\psi \label{eq:vprime}\\
    v\gamma' &= c_Lv^2\cos\phi - \cos\gamma - w'\sin\gamma\sin\psi \label{eq:gammaprime}\\
    v\cos\gamma\psi' &= c_Lv^2\sin\phi  + w'\cos\psi \label{eq:psiprime}\\
    \tilde{z}' &= v\sin\gamma\label{eq:zprime}\\ 
    w' &= \partial_{\tilde{z}} w \tilde{z}'\label{eq:wprime}\\
\tilde{x}' &= v\cos\gamma\cos\psi  \label{eq:xprime}\\
\tilde{y}' &=  v\cos\gamma\sin\psi  - w\label{eq:yprime}
  \end{align}
\end{subequations}

\subsection*{Two-Dimension limit}

The Rayleigh model is defined as follows: the wind field takes the simplified form $w(\tilde{z})\!=\!0$ for $\tilde{z}\!\le\! 0$ and $w(\tilde{z})\! =\! w_0$ for $\tilde{z}\!>\!0$; $w_0$ is the difference in wind speed between the two layers. 
$\tilde{z}<0$ represents the boundary layer, $\tilde{z}>0$ the wind layer, and $\tilde{z}=0$ an infinitely thin shear layer.
The time derivative  $w'$ in Eq.~(\ref{eq:wprime}) is zero nearly everywhere; the trajectory can therefore be decomposed into an equation of smooth evolution with $w'=0$,
 hereafter called ``glide'', and a discontinuous equation of layer transition where $w' = w_0 \delta$ with $\delta$ the Dirac distribution. In the remaining of this section we drop the subscript, and $w\ \hat{=}\ w_0$ is to be understood as the wind speed difference between the two layers.

Assuming that the glider trajectory is also nearly 2-D, the glide and transition equations take a simple form, given below.

\subsection*{Glide}

Consider the dynamics of a glider evolving according to  Eq.~(\ref{eq:eom}) in the vicinity $z = 0^\pm$ of the separating plane but not crossing the separation layer.
The 2-D approximation $z = 0^\pm$ brings $\gamma, \gamma' = 0$. Eq.~(\ref{eq:gammaprime}) becomes a constraint on the roll angle $\cos\phi = \frac{1}{c_Lv^2}$ and Eq.~(\ref{eq:eom}) simplifies to
\begin{subequations}
  \begin{align}
    v' &= -c_Dv^2 \label{eq:vdot}\\
    \psi' &= c_Lv\sin\phi \label{eq:psidot}
  \end{align}
\end{subequations}
Eliminating $\tau$, the parametric evolution of $v$ follows:
\begin{equation}
  \label{eq:dvdpsi}
  \frac{dv}{d\psi} = -\frac{1}{f}\frac{v}{\sqrt{1 - \frac{1}{c_L^2v^4}}}\cdot\sign(\psi')
\end{equation}
reflecting the airspeed cost of turning. The sign function is a consequence of the decrease of airspeed with time. Here, $f = c_L/c_D$ is the glide ratio.

\subsection*{Layer Transition}

During layer transition the forces remain finite, but $w' = w\delta$ induces a finite change of the glider's state. The state transition $(\psi_-, v_-)\mapsto(\psi_+, v_+)$ can be easily computed from groundspeed continuity (a consequence of the forces remaining finite). In airspeed quantities it translates to $\mathbf{V^+} = \mathbf{V^-} \pm W \mathbf{j}$ depending on whether the transition is up or down. This leads to
\begin{subequations}
\label{eq:trans}
  \begin{align}
  \tan \psi^+ &= \tan\psi^- \pm \frac{w}{v^-\cos\psi^-}\label{eq:transpsi}\\
   v^+ &= v^-\sqrt{1 \pm 2w/v^-\sin\psi^- + (w/v^-)^2}\label{eq:transV}
  \end{align}
\end{subequations}
Note that Eq.~(\ref{eq:transpsi}) is also smooth near $\psi = \pm\pi/2$.

\subsection*{Cycle Periodicity}

Both the layer transition and the glide equation are invariant by the transformation $(w, \psi) \mapsto (-w, -\psi)$. This can be seen as the consequence of the fact that the airspeed gain of flying upwind out of the boundary layer is equal to that of flying downwind into the boundary layer. This symmetry is particular to the Rayleigh problem: a finite thickness shear layer or a constraint on the average travel direction would break it. 

As a consequence the physical cycle [transition up$\to$wind layer glide$\to$ transition down$\to$boundary layer glide] can be subdivided into two equivalent sub-units [transition$\to$glide]$\to$[transition$\to$glide], expanded below:
\[
\dots(\psi^+, v^+)_{\raisebox{-1ex}{$\scriptstyle n-1$}}
\underset{\raisebox{-2ex}{$\scriptstyle \text{glide}$}}{\to}
\underbrace{
\left[
(\psi^-, v^-)_{\raisebox{-1ex}{$\scriptstyle n$}}
\underset{\raisebox{-2ex}{$\scriptstyle \text{transition}$}}{\to}
(\psi^+, v^+)_{\raisebox{-1ex}{$\scriptstyle n$}}
\underset{\raisebox{-2ex}{$\scriptstyle \text{glide}$}}{\to}
\ \right]
}_{\text{cycle }n}(\psi^-, v^-)_{\raisebox{-1ex}{$\scriptstyle n+1$}}
\dots
\]

In a stationary cycle, the airspeed is periodic $v_{n+1} = v_{n}$ and the heading angle is anti-periodic $\psi_{n+1} = -\psi_{n}$. Therefore the heading angle evolves by $\psi^-_{n+1} - \psi^+_n =  \psi^-_{n+1} + \psi^+_{n+1}$ over a glide phase.

\subsection*{Large Glide Ratio Limit}

Previous studies~\cite{Bower2011} have shown that 
the necessary wind speed $w$ tends to $0$ as the glide ratio (or \emph{finesse}) $f$ tends to $\infty$.
We therefore assume $f\gg 1$ and $w \ll 1$ and look for stationary solution of the cycle. The transition relation simplifies as follows: to the dominant order $\psi^+ = \psi^-\ \hat{=}\ \psi_0$ and the airspeed gain is $\Delta v = w/v^-\sin\psi_0$. The glide relation gives the approximated airspeed loss during each turn $\Delta v = -\frac{v}{f\alpha}\Delta\psi$ with $\alpha = \sqrt{1 - 1/c_L^2v^4}$. The anti-periodicity of $\psi_0$ implies $\Delta\psi = 2\psi_0$. Equating airspeed loss and gain brings the equation for the average airspeed
\begin{equation}
\frac{v}{\sqrt{1 - 1/c_L^2 v^4}} = \frac{\sin\psi_0}{\psi_0}\frac{f}{2}w
\label{eq:DSfundamental}
\end{equation}
The minimum $w, v$ pair
\begin{equation}
  w^* = \frac{3^{3/4}\sqrt{2}}{c_L^{3/2}/c_D}\ ,\ \ 
  v^* = 3^{1/4}/\sqrt{c_L}
\label{eq:vwstar}
\end{equation}
is attained at the maximum of $\sinc\psi_0 = \sinc(0) = 1$.  Besides, $\psi_0 = 0$ also maximizes airspeed for a given $w$: the small turn trajectory is optimal both for maintaining airborneness in small winds and for maximizing airspeed in large winds.
Note that as long as the approximation $f\gg 1$ holds, the glider's aerodynamic performance measure for the Rayleigh problem is, perhaps unsurprisingly, the minimum power coefficient $c_L^{3/2}/c_D$.

\subsection*{General Case}

Assuming either constant $c_L$, or optimal $c_L(v)$ to minimize the airspeed loss, Eq.~(\ref{eq:dvdpsi}) can be formally integrated by separation of variables and the exact solution to the Rayleigh problem becomes a relatively simple nonlinear algebraic problem.


\section*{ST2 Numerical solution by direct collocation}

\subsection*{Numerical procedure}

Our numerical model for Fig.~\ref{fig:numerical} and \ref{fig:2dpredict} is based on the \textsc{eom} of Eq.~\ref{eq:eom} with $w(z) = \frac{w_0}{1 + \exp-z/\delta}$. We formally rewrite the \textsc{eom} $\dot{\x} = f(\x, \uu)$. The question that we want answered is the following:
\emph{For a given glider $(c_{D,0}, f_{\max})$ and a given shear thickness $\delta$, what is the minimum wind amplitude $w_0$ that has feasible trajectories, periodic in the state $\x$?}
More specifically, for the traveling trajectories (right-hand side of Fig.~\ref{fig:numerical} and Fig.~\ref{fig:2dpredict}), the boundary conditions are 
$V(T) = V(0), \psi(T) = \psi(0), \gamma(T) = \gamma(0), z(T) = z(0)$. 
For the circular trajectories (left-hand side of Fig.~\ref{fig:numerical}, we imposed the boundary conditions $V(T) = V(0), \psi(T) = \psi(0) + 2\pi, \gamma(T) = \gamma(0), z(T) = z(0), x(T) = x(0)$. Note that the $x$-constraint in the latter set of boundary conditions is not strictly required. Without it the upper half cycle tends to peak at a higher altitude, with very small airspeed and very large $c_L$. The $x$-constraint  maintains $c_L$ to realistic values while conserving the main features of the unconstrained trajectories.

The question is cast into a finite dimensional optimization problem by direct collocation. First, time over one period $T$ is discretized into timesteps $[0, n_1 T, n_2 T, \dots, n_{N-1} T, T]$ with $0<n_1<\dots<n_{N-1}<1$. The spacing need not be uniform. We use the shorthand $\x_i \hat{=} \x(n_i T), \uu_i = \uu(n_i T)$. Following \emph{e.g.}~\cite{Flanzer2012,Hargraves1987}, the continuous-time constraints $\x(n_i T) = \int_{n_{i-1}T}^{n_i T}f(\x(t), \uu(t)) dt$ are approximated by
\[
\begin{aligned}
\uu_{m_i} &= \frac{1}{2}(\uu_i + \uu_{i-1})\\
  \x_{m_i} &= \frac{1}{2}(\x_i + \x_{i-1}) - \frac{1}{8}(f(\x_i, \uu_i) - f(\x_{i-1}, \uu_{i-1}))(n_i - n_{i-1})T\\
  0  &= \mathbf{C}_i = \x_{i-1} + \frac{1}{6}\left(f(\x_i, \uu_i) + 4 f(\x_{m_i}, \uu_{m_i})
    + f(\x_{i-1}, \uu_{i - 1})\right)(n_i - n_{i-1})T\\
\end{aligned}
\]


For the traveling problem, the previous discretization leads to the following nonlinear program (NLP):
\begin{equation}
  \label{eq:travel_dircol}
  \begin{aligned}
    \underset{\x_0,\dots,\x_N, \uu_0,\dots,\uu_N,w_0,T}{\text{minimize}}\quad & w_0\\
    \text{subject to}\quad& \mathbf{C}_i = 0,\quad i=1,\dots,N\\
  & (V_N, \psi_N, \gamma_N, z_N) =  (V_0, \psi_0, \gamma_0, z_0)\\
\text{and}\quad &z_0 = 0\\
& V_i, c_{L,i} >0\\
&-\pi <\psi_i< \pi,\ -\pi/2 < \gamma_i<\pi/2\\
  \end{aligned}
\end{equation}
A solution to the NLP is a feasible trajectory that locally minimizes the wind required for flight. Note that the last three relations are purely technical and the inequalities constraints were not active upon solution convergence.

Similarly, the circular problem is cast into
\begin{equation}
  \label{eq:circ_dircol}
  \begin{aligned}
    \underset{\x_0,\dots,\x_N, \uu_0,\dots,\uu_N,w_0,T}{\text{minimize}}\quad & w_0\\
    \text{subject to}\quad& \mathbf{C}_i = 0,\quad i=1,\dots,N\\
  & (V_N, \psi_N, \gamma_N, z_N, x_N) =  (V_0, \psi_0 + 2\pi, \gamma_0, z_0, x_0)\\
\text{and}\quad &z_0 = 0\\
& V_i, c_{L,i} >0\\
&-3\pi <\psi_i< 3\pi,\ -\pi/2 < \gamma_i<\pi/2\\
  \end{aligned}
\end{equation}

The problem was then solved for various $(c_{D, 0}, f_{\max}, \delta)$ with a nonlinear solver \emph{e.g.} SNOPT. We typically used $N=140$ time steps, leading to $O(1000)$ variables and constraints. Our Python implementation converged in $O(1-10)$ minutes on a 2013 Macbook Pro. We used more timesteps than in similar studies. The main reason for this choice is that for small $\delta$ the transition through the shear layer is of short duration, and resolving it requires a high level or granularity. To reach very small values of $\delta$ and validate the convergence  of our numerical model to our analytic model, we leveraged on the possibility to utilize non-uniform time spacing: we started by solving problems with large $\delta$ and subsequently adressed smaller $\delta$ by adaptively refining the time spacing near the transition in order to maintain a sufficient resolution.

\subsection*{Results}

The raw results for the cases illustrated in Fig.~\ref{fig:numerical} are collected in Figs. ~\ref{fig:num_raw_2}--\ref{fig:num_raw_2048} and \ref{fig:num_raw_2_circ}--\ref{fig:num_raw_2048_circ}. For case with $\delta = 1/128$ and  1/2048, the control points are non-uniformly spaced and are more dense near the transtion $z=0$. For both the circular and traveling cases, $\delta = 1/128$ and $1/2048$ are qualitatively similar. The boundary thickness for the albatross is closest to case $\delta = 1/128$. For the traveling cases $\delta = 1/128$ and 1/2048, the sub-periodicity discussed in ST1 is visible -- a qualitative difference from $\delta = 2$. In contrast, all circular cases are qualitatively similar to each other accross the range of $\delta$'s.

The main characteristics of the numerical model are collected in Figs.~\ref{fig:2dpredict} and \ref{fig:2dpredict_aux}.


\section*{ST3 Dimensions for the albatross flight}

Our numerical procedure solved the non-dimensional equations of DS. Fig.~\ref{fig:2dpredict} was dimensionalized with the typical characteristics of the wandering albatross as used in~\cite{Sachs2005} and reproduced in Table~\ref{tab:albatross}.
\begin{table}
  \centering
  \begin{tabular}{|lr|}
\hline
    Mass $m$ (kg) & 8.5\\
    Wing area $S$ (m$^2$) & 0.65\\
    Glide ratio $f_{\max}$ & 20\\
    Lift coefficient at $f_{\max}$, $c_{L, f_{\max}}$& 0.5\\
    $\lambda$ (m) & 21.8\\
    $V_c$ (m/s)& 14.6\\
\hline
  \end{tabular}
  \caption{Characteristic of the albatross used in this study. $\lambda$ and $V_c$ are calculated with the air density $\rho=1.2$~kg/m$^3$ and acceleration of gravity $g=9.8$~m/s$^2$.}
  \label{tab:albatross}
\end{table}

\begin{figure}
  \centering
  \includegraphics[width=\textwidth]{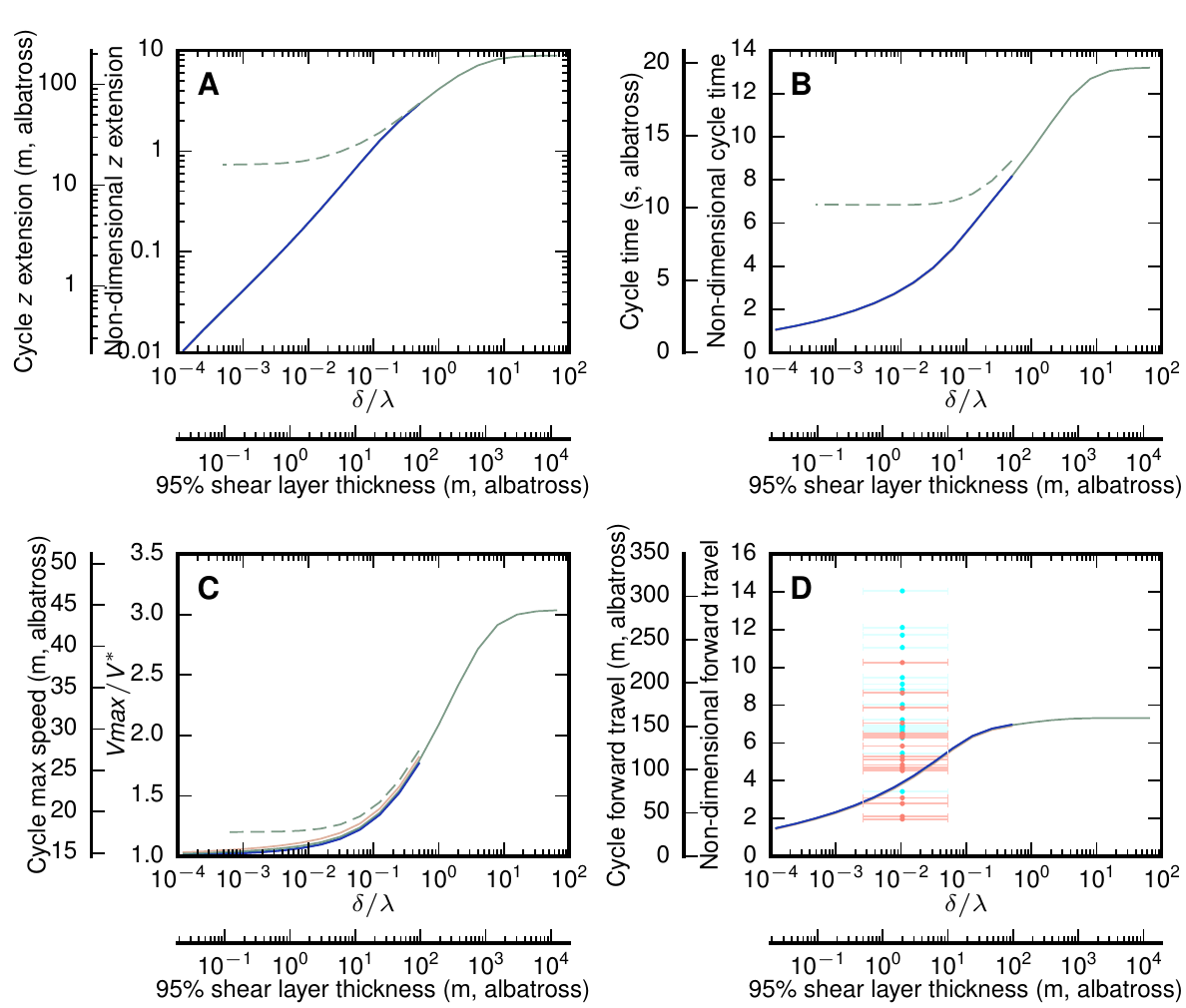}
  \caption{\textbf{Characteristics of the minimum-wind cycle} from ST2. Same legend as in Fig.~\ref{fig:2dpredict}.  (\textbf{A}) Height separation between the lowest and highest point of the cycle. For thin shears the traveling trajectory is nearly 2D. Note that the convergence rate is only about $z\sim \delta^{2/3}$. (\textbf{B}) Cycle duration. (\textbf{C}) Maximum airspeed attained during the cycle. (\textbf{D}) Forward travel during one cycle. The orange (\emph{resp.} cyan) dots correspond to twice the length of the sail (\emph{resp.} keel) phase in Fig.~\ref{fig:synthesis}A.}
  \label{fig:2dpredict_aux}
\end{figure}

As mentioned in the main text, the exact structure of the wind profile is one of the main sources of uncertainty in DS, and the boundary layer thickness parameter $\delta$ of our logistic model that best represents the wind experienced the albatross is similarly uncertain. 
A very reasonable assumption is that the effective shear layer thickness perceived by the albatross must be at least of the order of vertical extension of the albatross, from wingtip to wingtip, when it is in a bank. The albatross' span being $\sim$3~m, the shear layer perceived by the albatross must be thicker than $\sim$1~m. Pennycuick's description of the albatross performing ``swoops'' at the interface between windy and separated regions behind waves suggests that the actual thickness of the shear layer is ``small'', and that the effective thickness is $O$(1~m). Conversely, when the waves are small and the wind flow remains attached to the surface, it is possible that the albatross does not have access to the extremely thin boundary layer and as a consequence perceives a virtually thicker shear layer. The published data recording the vertical travel of the albatross is extremely scarce. The few data points in \cite{Sachs2012,Sachs2016} suggest that it is in the 5-15m range. Referring to Fig.~\ref{fig:2dpredict_aux}, such vertical travels correspond to a shear layer thickness of about 3m and 10~m respectively. Therefore, considerations on the albatross size as well as its reported vertical travel suggest to a shear layer thickness of the order of 1 to 3 meters, and at the very maximum 10~m. In Fig.~\ref{fig:2dpredict}B, 
we used the median hypothesis of 2~m with an error bar ranging from 1/2~m to 10~m.

Note also that the $n\%$ shear layer thickness is defined as the thickness, centered around 0, over which the wind field changes by $n\%$, \emph{i.e.} the height difference $z_{\min} - z_{\max}$ such that $W(z_{\min})/W_0 = \frac{n}{2}\%$ and $W(z_{\min})/W_0 = (1 - \frac{n}{2})\%$.

 The actual wind during flight of Fig.~\ref{fig:synthesis}A is reported in~\cite{Sachs2016} to be 7.8~m/s. In Fig.~\ref{fig:2dpredict}A, our estimate of the wind intensity $W_0$ perceived by the albatross is smaller: as discussed in the main text, the albatross is only able to harvest a fraction of that wind difference. Indeed even behind separated waves, the mass of air is not at rest with respect to Earth but typically travels at \emph{e.g.} the wave phase speed~\cite{Gent1977}. In non-separated flows the wind at 1m is typically more than 50\% the wind at 10~m and here again, the albatross can only exploit a fraction of the total wind speed \cite{Flanzer2012}. In the present study we assume that the albatross may access 25 to 50\% of the reported wind speed at 10~m.

\section*{ST4 \textsc{gps} data analysis}

\begin{figure}
  \centering
  \includegraphics[width=4.75in]{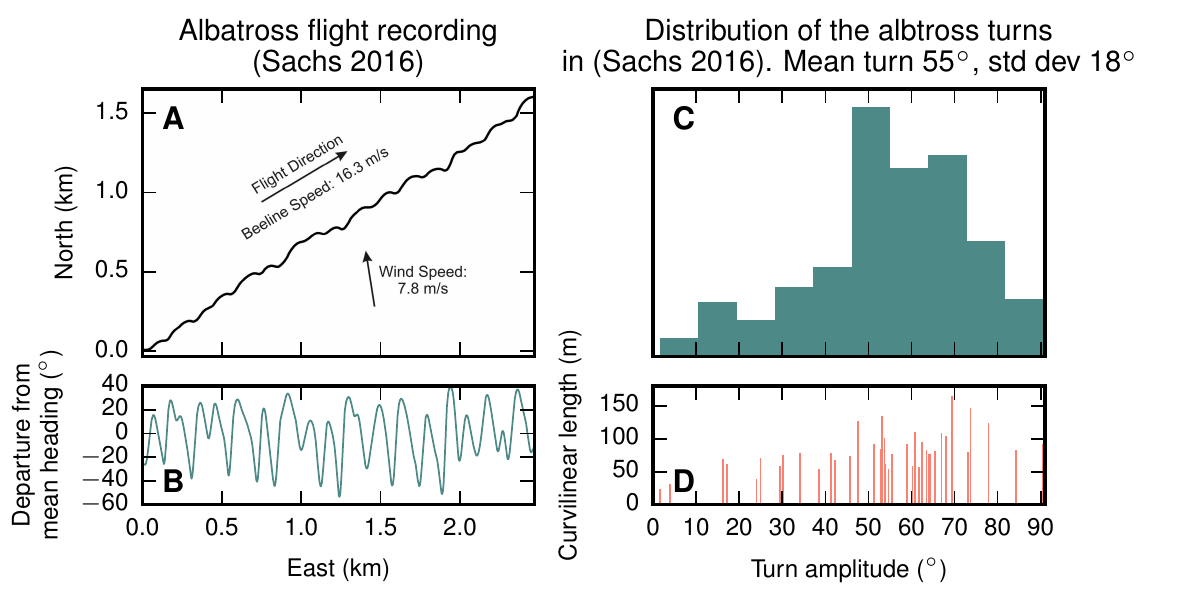}
  \caption{\textbf{Analysis of the albatross' trajectory}. (\textbf{A}) Recording of an albatross travelling accross a low wind \cite{Sachs2016}. (\textbf{B}) Albatross heading along the trajectory. In (\textbf{C}) the statistical analysis of the flight shows that the albatross turns on average by 55\degree{} (median also 55\degree{}), significantly less than the 180\degree{} of the half-turn picture. In this particular recording, the albatross virtually never turns more than 90\degree{}. (\textbf{D}) Curvilinear length of the individual turns.}
  \label{fig:sachs2016}
\end{figure}

\begin{figure}
  \centering
  \includegraphics[width=\textwidth]{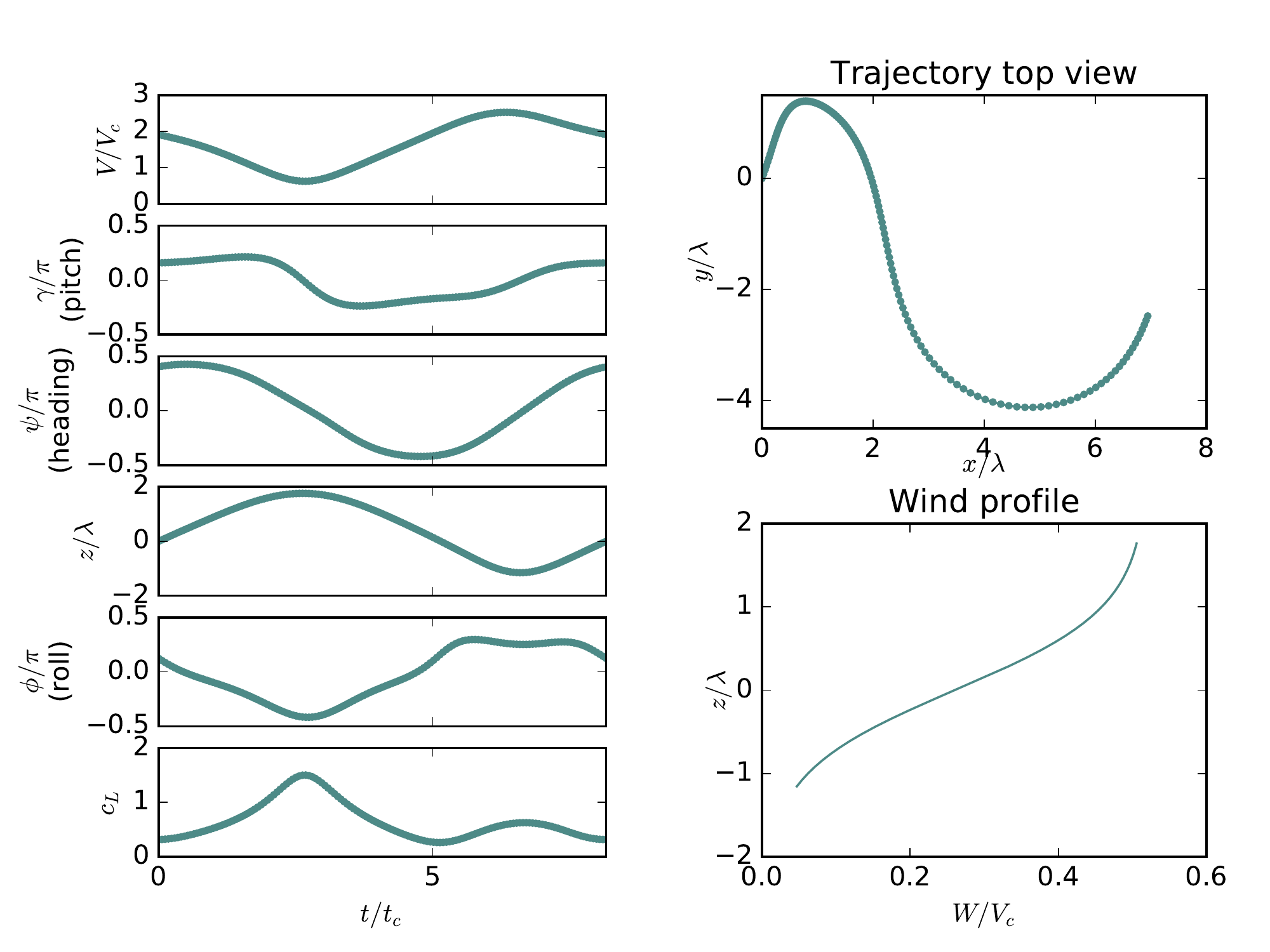}
  \caption{\textbf{Solution to the Rayleight problem} for $f_{\max} = 20, c_{L, f_{\max}} = 0.5, \delta = \lambda/2.$ $w_0 = 0.52$.}
  \label{fig:num_raw_2}
\end{figure}

\begin{figure}
  \centering
  \includegraphics[width=\textwidth]{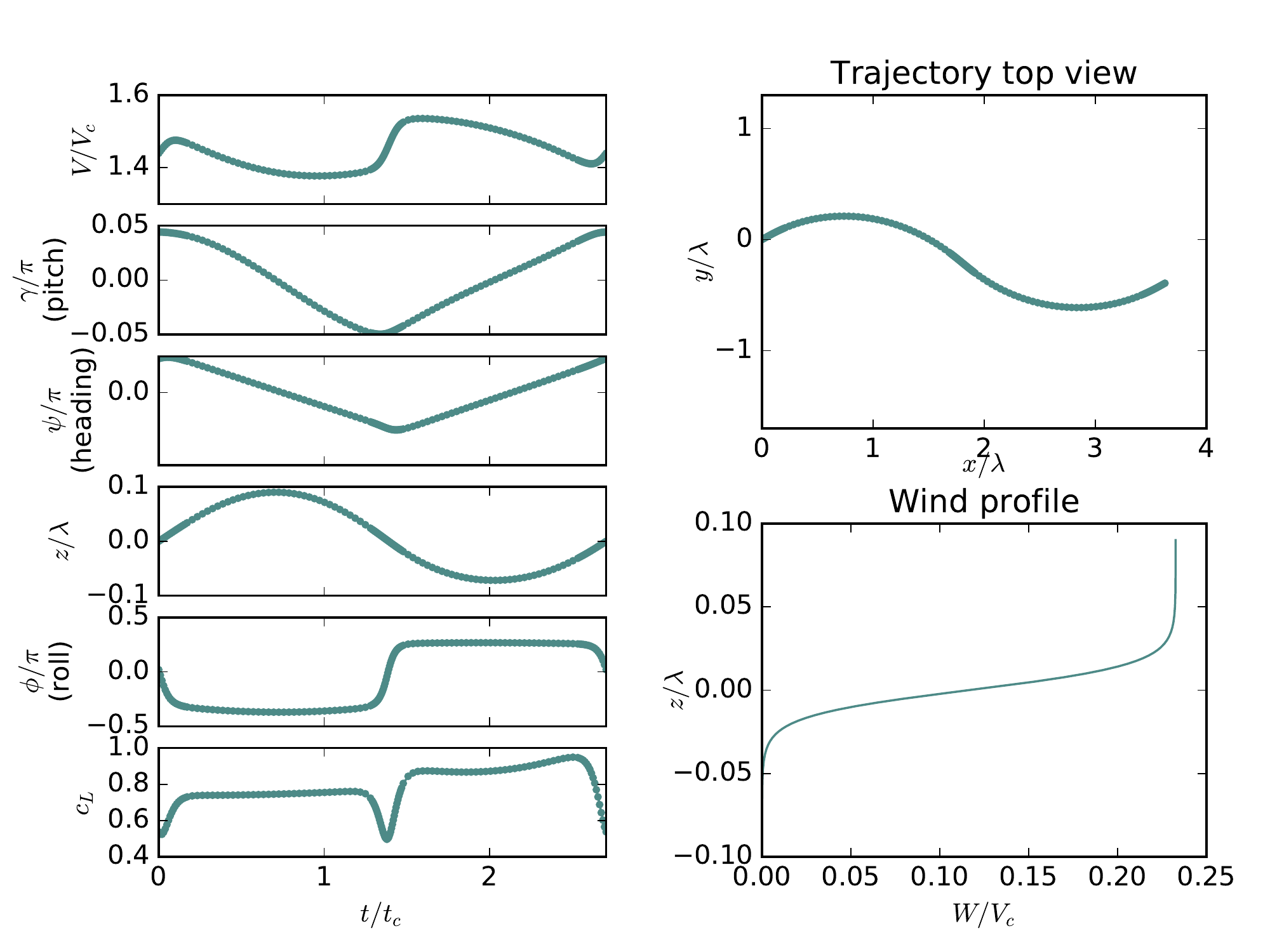}
  \caption{\textbf{Solution to the Rayleigh problem} for $f_{\max} = 20, c_{L, f_{\max}} = 0.5, \delta = \lambda/128$. $w_0=0.23$.}
  \label{fig:num_raw_128}
\end{figure}

\begin{figure}
  \centering
  \includegraphics[width=\textwidth]{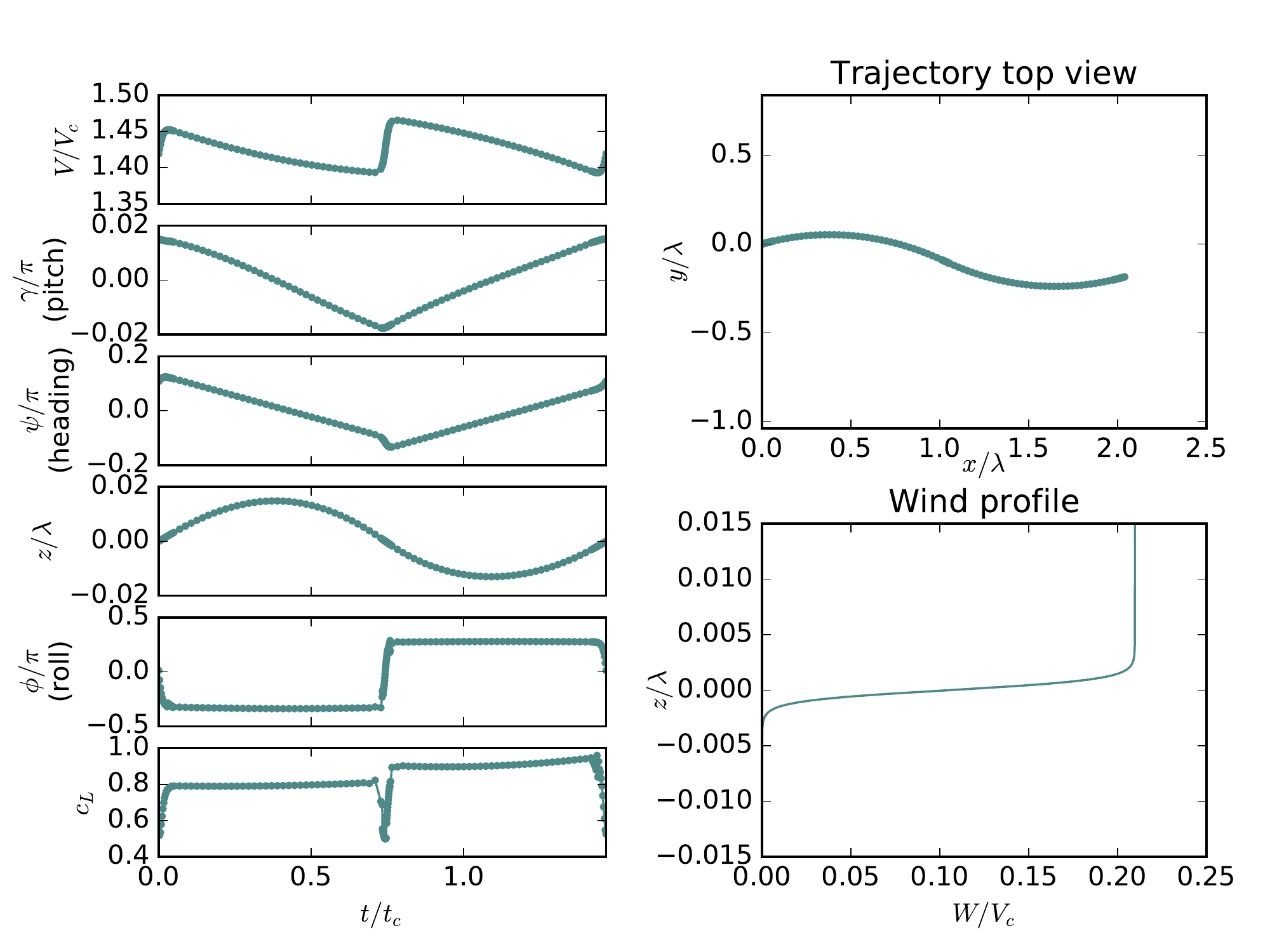}
  \caption{\textbf{Solution to the Rayleigh problem} for $f_{\max} = 20, c_{L, f_{\max}} = 0.5, \delta = \lambda/2048$. $w_0 = 0.21$.}
  \label{fig:num_raw_2048}
\end{figure}

\begin{figure}
  \centering
  \includegraphics[width=\textwidth]{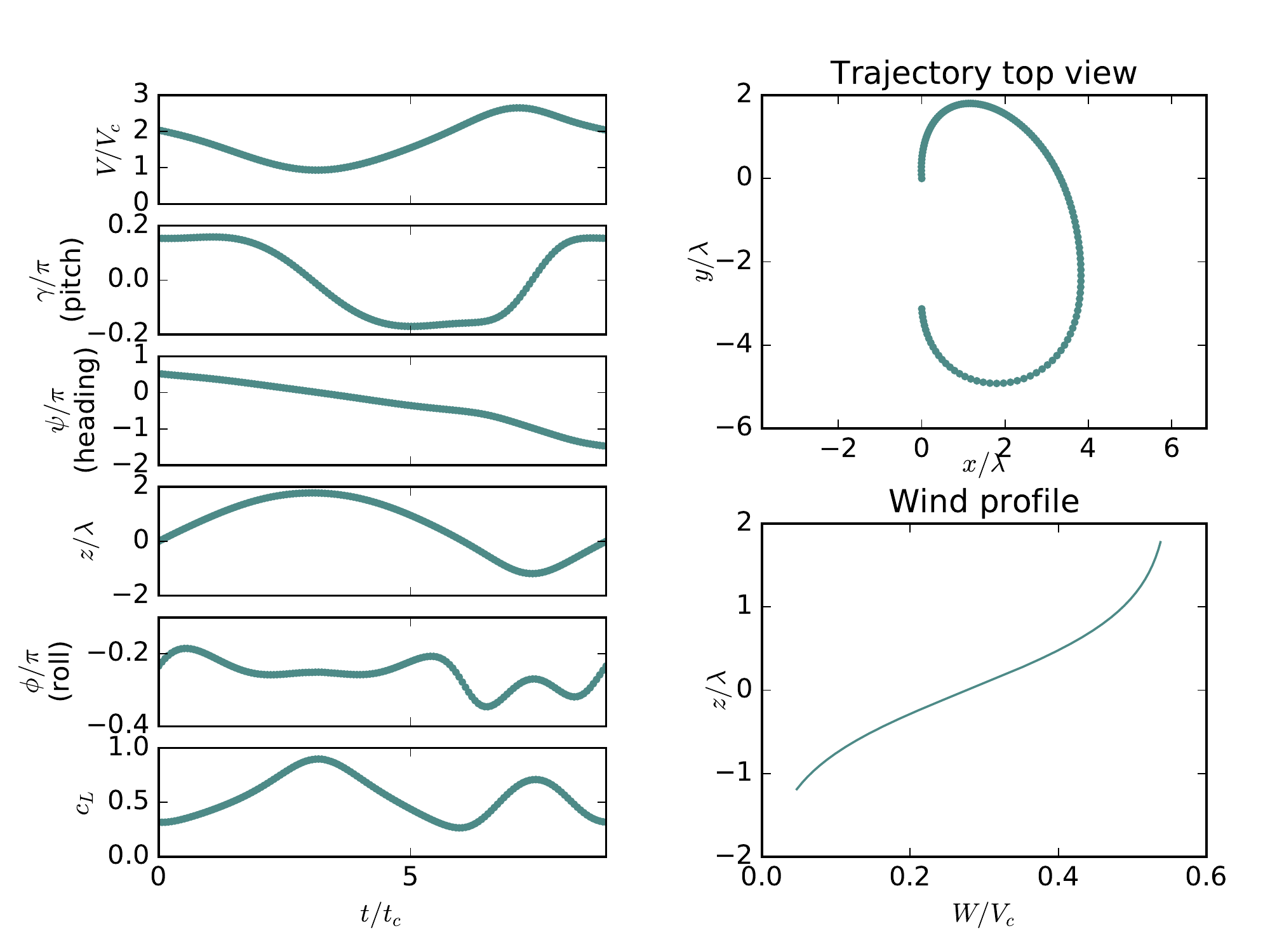}
  \caption{\textbf{Solution to the Rayleigh problem} for $f_{\max} = 20, c_{L, f_{\max}} = 0.5, \delta = \lambda/2$. $w_0 = 0.55$.}
  \label{fig:num_raw_2_circ}
\end{figure}

\begin{figure}
  \centering
  \includegraphics[width=\textwidth]{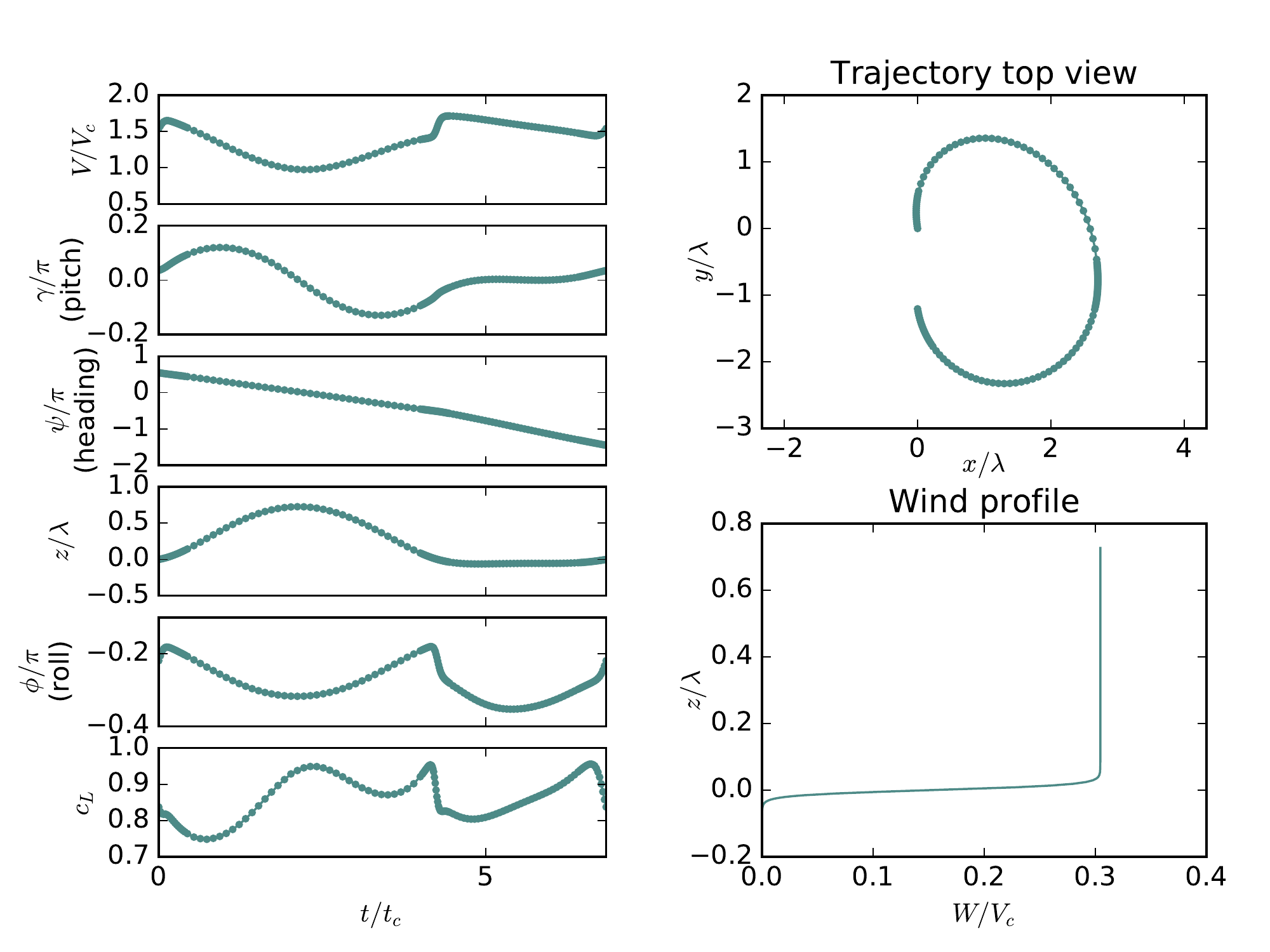}
  \caption{\textbf{Solution to the Rayleigh problem} for $f_{\max} = 20, c_{L, f_{\max}} = 0.5, \delta = \lambda/128$, $w_0=0.304$.}
  \label{fig:num_raw_128_circ}
\end{figure}

\begin{figure}
  \centering
  \includegraphics[width=\textwidth]{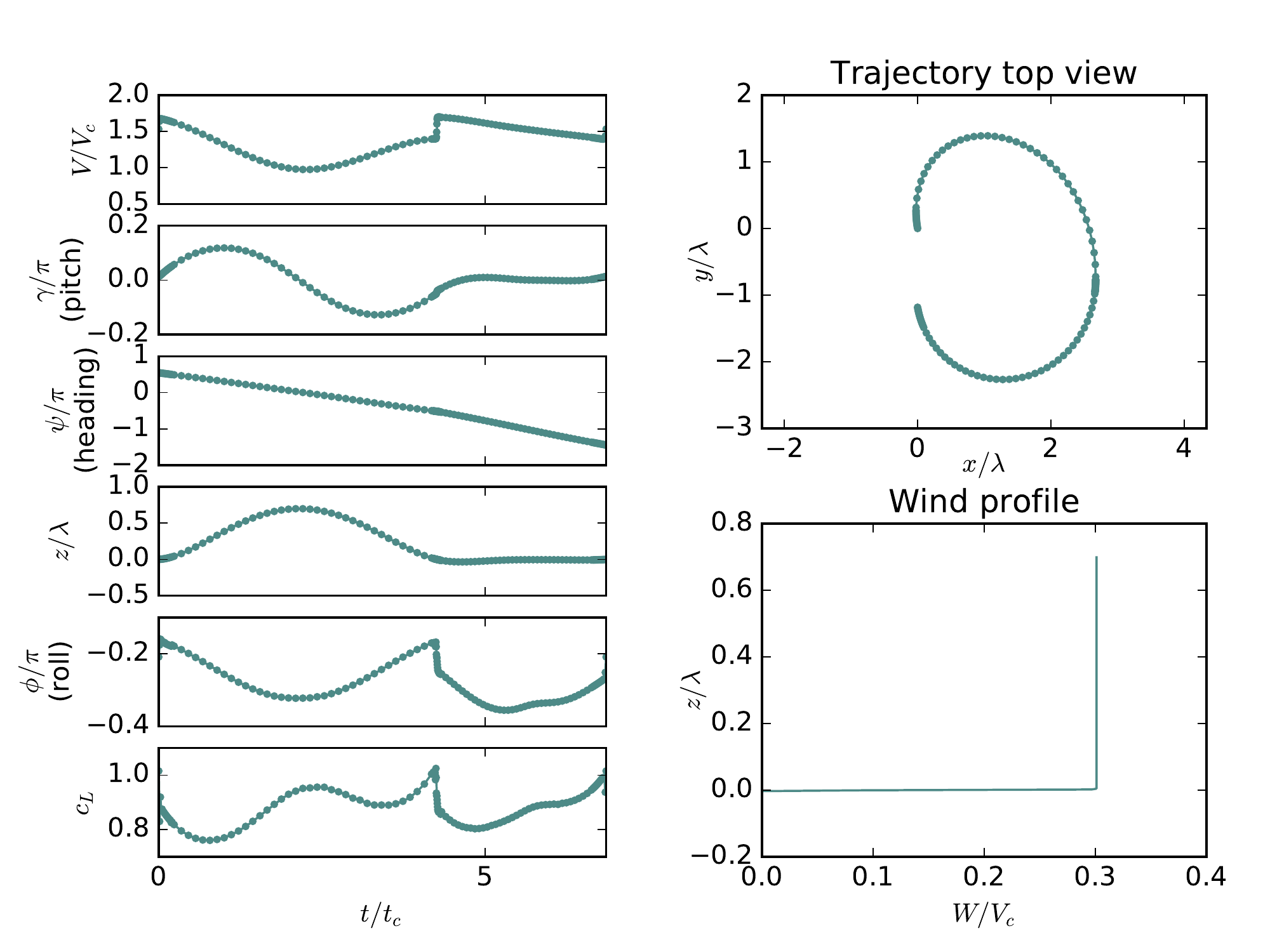}
  \caption{\textbf{Solution to the Rayleigh problem for} $f_{\max} = 20, c_{L, f_{\max}} = 0.5, \delta = \lambda/2048$, $w_0 = 0.301$.}
  \label{fig:num_raw_2048_circ}
\end{figure}

The data for Figs.~\ref{fig:synthesis} and \ref{fig:2dpredict} were extracted from the bitmap Fig.~11 of \cite{Sachs2016} (reproduced in Fig.~\ref{fig:sachs2016}A). For each pixel in the East direction, the center of the trajectory line was determined by an average operation. The result was filtered with the \texttt{filtfilt} filter from \texttt{scipy.signal}. The (ground) heading angle was then calculated (Fig.~\ref{fig:sachs2016}B). Figs.~\ref{fig:sachs2016}C and D report the distribution of the turns in the recording. The amplitude of each turn (in degrees) is also overlaid in Fig.~\ref{fig:2dpredict}B, with color darkness proportional to the curvilinear length of the turn (in meters). Similarly, the beeline progress is overlaid in Fig.~\ref{fig:2dpredict_aux}D. 

While our model predicts the albatross' turn amplitude extremely well, it underpredicts the cycle length (Fig.~\ref{fig:2dpredict_aux}D). Two factors may explain this: 1) There is a benefit in remaining in the  ``keel'' phase of the cycle because the ground effect reduces drag, possible uplift from wave-generated winds are a secondary source or energy, and the direction of travel is skewed upwind if more time is spent in the slow layer. This hypothesis is consistent with the observation that the keel phase is indeed typically longer than the sail phase. 2) In conditions of sufficient wind, secondary goals of the  albatross may include forward travel speed, minimum control activity and/or aerodynamic loads and reaching higher altitudes for \emph{e.g.} a better observation of the ocean. Those goals are better satisfied in long cycles. They also tend to skew the trajectories to smaller turns, consistent with the small over-prediction of the turn amplitude by our model.

\end{document}